\documentclass[aps,floats,floatfix,showpacs,superscriptaddress]{revtex4}

\usepackage{epsfig}
\newcommand{\be}{\begin{equation}}
\newcommand{\ee}{\end{equation}}
\newcommand{\bea}{\begin{eqnarray}}
\newcommand{\eea}{\end{eqnarray}}
\begin{document}

\title{Ordering phase transition in the one-dimensional Axelrod model}

\author{Daniele Vilone}
\affiliation{Dipartimento di Fisica, Universit\`a di 
Roma ``La Sapienza'', P.le A. Moro 2, I-00185 Roma, Italy}
\author{Alessandro Vespignani}
\affiliation{The Abdus Salam International Centre for Theoretical Physics
(ICTP), P.O. Box 586, 34100 Trieste, Italy}
\author{Claudio Castellano}
\affiliation{Dipartimento di Fisica, Universit\`a di 
Roma ``La Sapienza'', P.le A. Moro 2, I-00185 Roma, Italy}
\affiliation{INFM, Unit\`a di Roma 1, P.le A. Moro 2, I-00185 Roma, Italy}

\date{\today}
\begin{abstract}
We study the one-dimensional behavior of a cellular
automaton aimed at the description of the formation and evolution
of cultural domains.
The model exhibits a non-equilibrium
transition between a phase with all the system sharing the same culture
and a disordered phase of coexisting regions with different cultural
features.
Depending on the initial distribution of the disorder the transition
occurs at different values of the model parameters.
This phenomenology is qualitatively captured by a mean-field approach,
which maps the dynamics into a multi-species reaction-diffusion problem.
\end{abstract}
\maketitle

\section{Introduction}

In recent years cellular automata and lattice models
have been introduced in many fields in order to investigate the
emergence of nontrivial collective phenomena out of simple
microscopic interactions between a large number of elementary
objects~\cite{Marrobook,Jensenbook,Axelrodbook}.
In the context of social phenomena, numerical simulations of 
simple automata have been used to study the formation and spreading
of cultural domains~\cite{Axelrod97, Axtell96, Castellano00, Klemm02}.
Culture is intended here as the set of behaviors, beliefs,
technical standards or values that individuals may mutually exchange.
The cultural features are modeled as a set of $F$ variables,
which can assume integer values between 1 and $Q$, specifying
the various traits that each feature can assume~\cite{Axelrod97}.
In general, the environment where exchanges take place is a
two-dimensional square grid, where each site represents an individual 
or a group of individuals, whose culture is defined by the specific
values of the $F$ variables.
Starting from an initial random configuration,
interactions take place between nearest neighbors, leading to the
formation and evolution of regions of homogeneous culture.
It is obvious that the assessment of how realistic the model is or to
which extent its results can be useful in the interpretation of
empirical data in the social science context is a task left to sociologists.
From the point of view of statistical physics this model (Axelrod model)
is appealing, because it has a non-trivial out of equilibrium
dynamics, similar but not equivalent to other well studied
models~\cite{Liggett85, Dornic01}.

In previous works~\cite{Axelrod97,Axtell96,Castellano00,Klemm02}, 
it has been investigated how the model behavior changes with respect to
the values of the parameters $F$ and $Q$. In particular it has been shown by
large-scale simulations that, on a two-dimensional lattice, the model exhibits
a non-equilibrium transition between a phase with all
the system sharing a common culture and a disordered phase of coexisting
regions with different cultures.
This is a very interesting example of ordering phase transition in the context
of far from equilibrium statistical physics.
In this paper we perform a thorough analysis of the model defined in
Ref.~\cite{Axelrod97} on a one-dimensional lattice.
This geometry acquires a special role in the description of social
systems, since it can be used as starting point for complex topologies
such as small-world networks~\cite{Wattsbook,Strogatz01}.
Noticeably, in dimension $d=1$ it is possible to provide an analytical
treatment of the model by mapping the evolution dynamics into a multi-species
reaction-diffusion system, that we study within a mean-field approximation.
We recover also in one dimension the existence of an
ordering phase transition, occurring at a critical value of the parameter $Q$.
When the initial value of the cultural variables on each site is chosen
randomly with uniform distribution between 1 and $Q$, we find that the 
system undergoes a transition between a fully ordered state for $Q<Q_c$
and a disordered one for $Q>Q_c$.
When the distribution of initial variables is instead Poissonian of
parameter $q$, the system exhibits a phase transition between order
and disorder, but only for $F>2$.
For $F=2$ any finite value of $q$ leads, in the thermodynamic limit, to an
asymptotic disordered state.
In order to test the analytical predictions we perform numerical simulations
of the original dynamics, finding a good agreement with the theoretical
results.

The paper is organized as follows.
In Section~\ref{Sec2} the model is introduced, the dynamics qualitatively
described and the results of the two-dimensional case briefly summarized.
In Sec.~\ref{Sec4} we present the mapping to the reaction-diffusion system
and the mean-field approach to it.
In Sec.~\ref{Sec3} we report the results of numerical simulations
of the model in one dimension, both for the case of uniform and Poisson
initial distributions, and compare them with the analytical predictions.
Finally in Sec.~\ref{Sec5} we draw some conclusions.

\section{The model}
\label{Sec2}
We consider a one-dimensional lattice of linear size $L$.
On each site $i$ there are $F$ integer variables $\sigma_{i,f}$
which define the cultural features of the individuals living on that site.
In the original model~\cite{Axelrod97}, each feature $f=1,\ldots,F$
on each site $i$ is initially drawn randomly from
a uniform distribution on the integers between $1$ and $Q$.
The parameter $Q$ is a measure of the initial cultural variability
(i.e. disorder) in the system.
In Ref.~\cite{Castellano00} instead, the initial values were chosen
according to a discrete Poisson distribution  of parameter $q$
${\rm Prob}(\sigma_{i,f}=k)=q^k e^{-q}/k!$,
so that the positive real $q$ was the average of the values extracted. 
Results were found to be qualitatively similar independently
from the initial distribution, and only the Poissonian case was studied
in detail.
Here we will consider also the case with uniform distribution
and we will discuss analogies and difference between the two cases.

The dynamics is defined as follows.
At each time step, a pair of nearest neighbor sites
$i$ and $j$ is randomly chosen. A feature $f$ is selected
and if $\sigma_{i,f}\not =\sigma_{j,f}$ nothing happens.
If instead $\sigma_{i,f}=\sigma_{j,f}$ then an additional feature
$f'$ is randomly chosen among those taking different values across the
bond, $\sigma_{i,f'}\neq \sigma_{j,f'}$.
Such a feature is then set equal in the two sites:
$\sigma_{i,f'}\to \sigma_{i,f'}'=\sigma_{j,f'}$.
Time is measured as the total number of attempts divided by the number of
sites $L$.
Axelrod model can be seen as $F$ coupled voter models~\cite{Liggett85}.
The evolution of the system is basically a coarsening dynamics of ordered
regions. The initial state is totally disordered since the choice of
variables in each site is uncorrelated to any other one.
As the dynamics starts, the interaction between sites
tends to make neighbors similar, introducing correlations
in the system. Ordered regions start to form with all sites sharing
exactly the same value of all variables.
Clearly there are many different possible types of ordered regions.
As time proceeds boundaries between regions move, and the
average domain size increases.
On a finite lattice the ordering process continues until the system
reaches an absorbing state such that no further dynamics is possible.
Absorbing states can be of two types.
A perfectly ordered absorbing state is made by a single region
covering the whole system: all sites have the same set of variables
$\sigma_{i,f}$. Otherwise an absorbing state can also be made by
different ordered regions such that adjacent regions have all
variables different. In that case all sites are either totally 
equal or totally different from their neighbors and no interaction may take
place.

The nature of the final absorbing state depends on the full evolution
of the system, i. e. it is the result of the competition between the
disorder of the initial configuration and the drive toward order
due to local interactions.
Intuitively one expects that for small initial disorder (small $Q$ or $q$) 
local interactions will prevail and 
the final state will be perfectly uniform, while in the opposite
limit the ordering process will not last long enough to fully overcome
the disorder of the initial condition.
This picture turns out to be correct in two dimensions~\cite{Castellano00}.
In that case extensive numerical simulations have shown that
for small values of $q$, i. e. small initial disorder, the model
converges toward an ordered state where a single culture occupies
a macroscopically large fraction of the system. On the contrary
for large values of $q$ many different cultures coexist in the
final absorbing state. In the thermodynamic limit of an infinitely
large system one sees a well defined non-equilibrium phase transition
for a precise value $q_c$, separating an ordered phase (for $q < q_c$)
from a disordered one (for $q>q_c$).
Interestingly, the nature of the transition is different depending on the
number $F$ of variables used to define the culture of a site.
If $F=2$ the transition is continuous, with the order parameter (defined
below) vanishing for $q \to q_c$. For $F>2$ instead the order parameter
exhibits a jump at the transition. The different behavior is reflected also
in the distribution of region sizes at the transition. For $F=2$ it is
a power law with an exponent $\tau<2$, while for $F>2$ the exponent is
universal and larger than 2, so that the transition does not involve the
divergence of a correlation length~\cite{Castellano00}.
It is important to stress that the kind of nonequilibrium phase transition
discussed here is not an absorbing phase transition~\cite{Marrobook}.
Here the system always reaches a frozen absorbing state. The transition
occurs in the properties of such absorbing state and hence is static
in nature, although it is clearly determined by the dynamical evolution
of the problem.

In this paper we shall investigate whether the same kind
of transition occurs also in a one-dimensional geometry.
To monitor the evolution of the model we mainly focus on the density
(i. e. the number divided by the system size $L$) of active bonds
$n_A(t)$ and of {\em fixed} bonds $n_F(t)$.
We define as {\em active} a bond  between two adjacent sites such
that they have more than 0 and less than $F$ equal features.
We define as {\em fixed} a bond between two adjacent sites such
that they have no equal variables.
Clearly, active bonds are where the dynamics take place and form
the boundary between regions that may interact, leading to further
ordering. Boundaries formed by fixed bonds are instead frozen.
All bonds connecting sites belonging to the same region are
neither active nor fixed. We denote their density as $n_0$. Clearly,
for all times $n_0+n_A+n_F=1$.
Both the density $n_A$ of active bonds and the density $n_F$ of fixed
bonds measure correlations in the system: $1/n_F$ and $1/n_A$
are correlation lengths growing in time.
Since for any finite lattice an absorbing state is always reached by the
dynamics, $n_A(t)$ goes asymptotically to zero in all cases.
The value of $n_F(t=\infty)$ measures the order in the system.
If $n_F(\infty)$ keeps finite as $L$ diverges, the system remains disordered
in the thermodynamic limit, while $n_F(\infty) \to 0$ indicates full order.

Another useful quantity that will be considered below, is
$\langle S_{max}\rangle/L$, where
$\langle S_{max} \rangle$ is the average over different realizations of the
size of the largest region existing in the final
frozen state~\cite{Castellano00}.
$\langle S_{max} \rangle /L$ is the fraction of the whole lattice occupied
by the largest region and is an alternative order parameter,
being zero for a disordered state and finite when a
macroscopically large region exists in the system.

In the next Section we describe an analytical treatment of the problem
based on the mapping of the dynamics into a multi-species
reaction-diffusion system, which is then analyzed in terms of mean-field
equations. This approach turns out to capture remarkably well the
qualitative behavior of the system. In some cases
it gives predictions that are quantitatively accurate.

\section{Theoretical approach}
\label{Sec4}

\subsection{Mapping to a reaction-diffusion system}

The Axelrod model considers the sites as elementary objects, which interact
according to the rules described above.
This dynamics can also be thought in terms of bonds between sites
whose dynamics is induced by the updating of sites.
More precisely, let us define as bond of type $k$ a bond connecting
two sites with $k$ different features.
Clearly $k$ can be any number between $0$ and $F$.
Bonds of type $0$ connect sites with all variables equal, i. e. belonging
to the same region.
Bonds of type $F$ instead connect sites completely different and are
therefore the fixed bonds introduced above.
Bonds of type $1, \ldots, F-1$ are what we called active bonds in the
previous section.

To understand how the update of sites induces a dynamics for bonds, let us
consider three neighbor sites (A), (B), (C).
\begin{figure}
\centerline{\psfig{figure=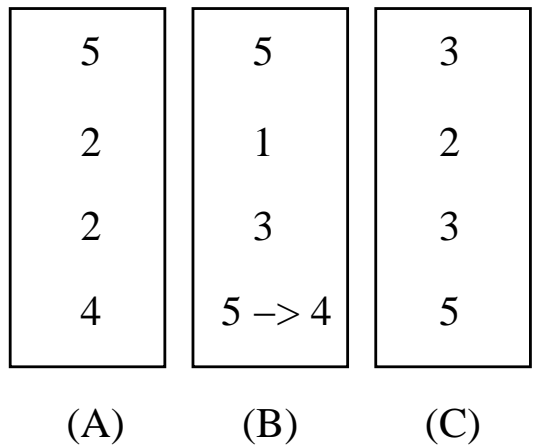,width=7cm,angle=0}}
\caption{Example of the evolution of three neighboring sites for $F=4$.
Site (B) and its neighbor (A) are selected and their feature $f=1$ is
compared. Since $\sigma_{A,1} = \sigma_{B,1} = 5$, the two sites interact
and feature $f'=4$ of site (B) is set equal to $\sigma_{A,4}=4$.
Bond (A)-(B), which was of type 3 becomes of type 2.
Bond (B)-(C), initially of type 2 becomes of type 3.}
\label{Fig1}
\end{figure}
Site (B) is updated depending on the configuration of its neighbor (A).
If the bond between sites (A) and (B) is of type $0$ or $F$ nothing
occurs. Otherwise, if the bond is of type $n$ one variable on site
(B) is set equal to the corresponding variable on site (A). Hence the
bond (A)-(B) becomes of type $n-1$.
The update of site (B) modifies in general also the state of bond (B)-(C),
which initially was of type $m$. Notice that for this bond the number
of features which are different may increase, decrease or even remain
unchanged, so that after the update the bond may be of type $m-1$,
$m$ or $m+1$.
In summary, if we denote a bond of type $k$ as $A_k$, the update
of a site is then equivalent to a two-particle reaction
\be
A_n + A_m \to  A_{n-1} + A_{m+j}~~~~~~~~~~(j=0,\pm 1).
\label{reactions}
\ee
A specific example is presented in Fig.~\ref{Fig1}.
Bonds of type $0$ can be considered as a sort of vacuum state where
other types of bonds move. Reactions of the type $A_1 + A_0 \to
A_0+ A_1$ are then diffusion processes for bonds of type $1$.
These are the only diffusion processes allowed.

In order to fully specify the dynamics we must provide the transition
rates for the reactions~(\ref{reactions}).
Let us consider the three sites of Fig.~\ref{Fig1}. 
One of the features of site (B) is selected and compared
with the corresponding one of site (A).
In order for a reaction to occur the two variables must be the same.
This occurs with a probability $R^{(1)}_n=(1-n/F)(1-\delta_{n,0})$
where we have taken into account that nothing happens if the bond is of
type $0$.

Hence with probability $R^{(1)}_n$, another variable $\sigma_{B,f'}$
is set equal to  $\sigma_{A,f'}$.
What is the effect of this change on the bond (B)-(C)?
If $\sigma_{C,f'}$ was equal to $\sigma_{B,f'}$, then the update increases
by one the number of features different in the bond (B)-(C).
The probability that $\sigma_{C,f'}=\sigma_{B,f'}$ is the probability
that we randomly select one of the $F-m$ equal variables out of the $F$
total variables
\be
R^{(2)}_{m,+1} = {F-m \over F} = 1-{m \over F}.
\ee
With probability $1-R^{(2)}_{m,-1}$ instead, the variable selected
will be different in site (B) and (C),  $\sigma_{C,f'} \neq \sigma_{B,f'}$.
In this case, since $\sigma_{B,f'}$ is set equal to $\sigma_{A,f'}$
the final state of bond (B)-(C) depends on whether $\sigma_{A,f'}$
and $\sigma_{C,f'}$ are equal or not. 
Let us denote as $\lambda$ the probability that $\sigma_{A,f'}=\sigma_{C,f'}$.
Then the probability that bond (B)-(C)
goes from type $m$ to type $m-1$ is
\be
R^{(2)}_{m,-1} = {m \lambda \over F}.
\ee
The probability $R^{(2)}_{m,0}$ is obtained via the normalization condition
\be
R^{(2)}_{m,0} = {m (1-\lambda) \over F}.
\ee
In summary the reaction $A_n + A_m \to  A_{n-1} + A_{m+j}$
occurs with rate
\be
R^{(1)}_n R^{(2)}_{m,j}.
\label{rates}
\ee

The reactions~(\ref{reactions}) and the rates~(\ref{rates}) fully specify
the dynamics of the model. Provided one is able to write down the
appropriate form of $\lambda$ the mapping from the original Axelrod model
is exact and Eqs.~(\ref{reactions}) and~(\ref{rates}) are an alternative
formulation of the same dynamics.

\subsection{Mean-field approximation}

As discussed above, $\lambda$ is the probability that two independent random
variables (A and C), different from a third one (B), are equal.
Here we will determine it by introducing a mean-field approximation,
i. e. assuming spatial uniformity.

Let us call $k$ the value of variable (B) and $p_B(k)$ its distribution.
The $\lambda$ can be written
as $\sum_k p_B(k) \lambda_k$ where $\lambda_k$ is the probability 
that A and C assume the same value, different from $k$.
Clearly
\be
\lambda_k = \sum_{i \ne k, j \ne k}
{\tilde p}_A(i) {\tilde p}_C(j) \delta_{i,j}
\ee
where ${\tilde p}_A(i) = p_A(i)/\sum_{i \ne k} p(i)$ is the probability
distribution of the variable $A$ constrained to be different from $k$.
We assume that the distributions of all variables A, B and C remain
equal to the initial ones at all times.
For uniform initial distribution $p(i)=1/Q$, we recover the expected value
\be
\lambda={1 \over Q-1}.
\ee

\begin{figure}
\centerline{\psfig{figure=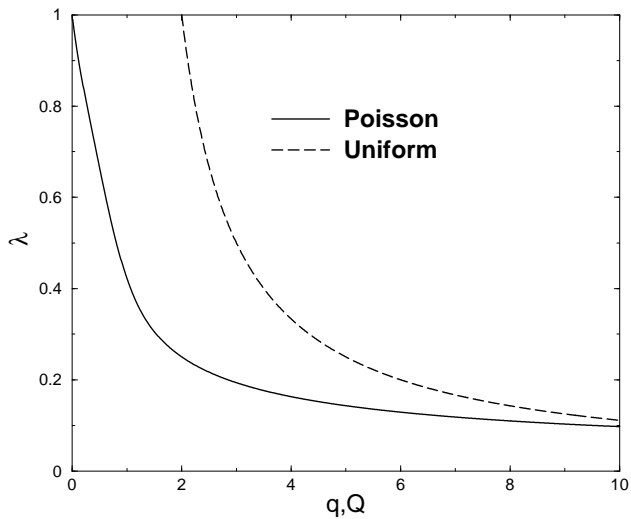,width=7cm,angle=-90}}
\caption{Plot of $\lambda$ as a function of $Q$ (uniform case) or $q$
(Poisson case)}
\label{Fig2} 
\end{figure}

For variables distributed according to a Poissonian,
the plot of $\lambda$ as a function of $q$, is presented in Fig.~\ref{Fig2}.
The two behaviors are qualitatively similar. However, one has to remark
that in the Poissonian case $\lambda=1$ for $q=0$, in the trivial situation
when all variables are the same. For uniform initial condition instead
$\lambda=1$ for $Q=2$, i. e. when the variables may assume {\em two} values.

From the knowledge of the possible reactions and their associated
transition rates it is straightforward to write down the mean-field
equations for the densities $n_k$ of bonds of type $k$.

We obtain the set of equations ($k=0,\ldots,F$)
\bea
\dot{n}_k &=& - R^{(1)}_k n_k + R^{(1)}_{k+1} n_{k+1} \\ \nonumber
&+& \sum_{N=1}^{F-1} R^{(1)}_N n_N 
\left[-n_k \left(1- R^{(2)}_{k,0} \right) +
n_{k-1} R^{(2)}_{k-1,1} + n_{k+1} R^{(2)}_{k+1,-1} \right]
\label{MFeq}
\eea
where clearly $n_{-1} \equiv 0 \equiv n_{F+1}$. More explicitly
\bea
\dot{n}_k &=& - \left(1-{k \over F} \right) (1-\delta_{k,0}) n_k 
+  \left(1-{k+1 \over F} \right) n_{k+1} \\ \nonumber
&+& \sum_{N=1}^{F-1}  \left(1-{N \over F} \right) n_N 
\left[-n_k \left(1- {k \over F} (1-\lambda) \right) +
n_{k-1} \left(1-{k-1 \over F} \right) + n_{k+1} {k+1 \over F} \lambda \right].
\eea
It is easily verified that $\sum_{k=0}^F \dot{n}_k=0$.
Eqs.~(\ref{MFeq}) are mean-field equations, since they
neglect spatial fluctuations and noise in the system.
They are complemented by the initial conditions, which are
\be
n_k(t=0) = P_{eq}^{F-k} \left(1-P_{eq}\right)^k \left(\begin{array}{c}
F \\
k
\end{array}
\right)
\ee
where $P_{eq}$ is the probability that a feature takes the same
value in two adjacent sites. Clearly $P_{eq} = 1/Q$ in the uniform case,
while
\be
P_{eq} = \sum_{h=0}^{\infty} {q^{2h} e^{-2q} \over (h!)^2}
\ee
for a Poisson initial distribution.

The mean-field system of equations of motion~(\ref{MFeq}) consists of
$F$ coupled nonlinear equations.
A detailed study of the temporal evolution is necessarily performed
by integrating them numerically.
However some important conclusions on the asymptotic state of the
system can be drawn analytically by considering the fixed points
of Eqs.~(\ref{MFeq}).
A first crucial observation is that, for any $F$, a state
with all $n_k=0$ ($k=1,\ldots,F-1$) is a fixed point of the equations.
Therefore the system always admits a family of frozen asymptotic states, 
parametrized by the density of fixed bonds $n_F$ ($n_0=1-n_F$ because
of normalization).
To understand which of these fixed points is selected by the dynamics
we perform a stability analysis, by linearizing the dynamical
system~(\ref{MFeq}) around the generic fixed point $n_k=0$.

\begin{figure}
\centerline{\psfig{figure=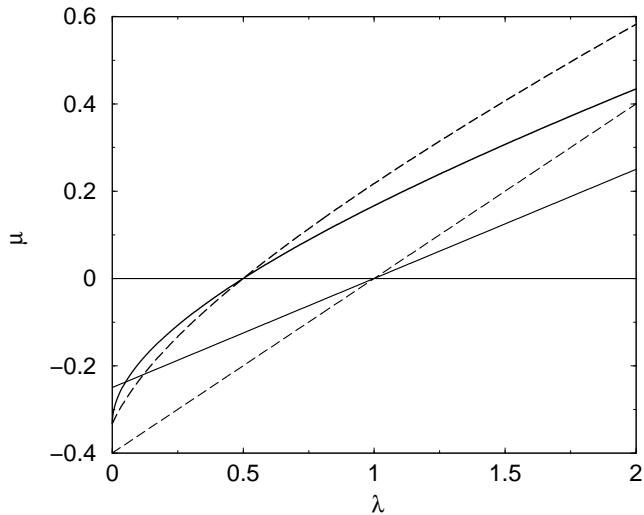,width=7cm,angle=-90}}
\caption{Plot of the eigenvalue $\mu$ versus $\lambda$ for $F=2$
(thin lines) and $F=3$ (thick lines), and $n_F=0.5$ (solid) and
$n_F=0.8$ (dashed).}
\label{Fig3}
\end{figure}

The stability matrix has $F$ eigenvalues.
One eigenvalue is always zero, corresponding to the
direction parametrized by $n_F$ and all made of fixed points.
$F-2$ other eigenvalues are real negative numbers or complex
ones with negative real part, corresponding to stable directions. 
The last one, that we denote as $\mu$, is always real and it crucially
depends on the value of $\lambda$~\cite{note}.
Its value as a function of $\lambda$, computed analytically for
$F=2$ and $F=3$, is plotted in Fig.~\ref{Fig3} for two different values
of $n_F$.
Clearly $\mu$ depends from the value of $n_F$, but for any $n_F>0$ it is
negative for $\lambda<1/(F-1)$ and positive for $\lambda>1/(F-1)$.
For $n_F=0$ (homogeneous absorbing state) $\mu=0$ for any value of
$\lambda$, so that the stability of this fixed point cannot be
assessed {\em directly} within a linear analysis.
However its stability can be inferred indirectly from the stability
of other fixed points: if all other fixed points are unstable the
fixed point with $n_F=0$ must be stable; if all others are stable,
the dynamical system will not go to $n_F=0$.
This allows us to conclude that for values of the parameter $Q$ or $q$ 
of the model such that $\lambda<1/(F-1)$ 
the equations will converge exponentially fast to a finite value $n_F>0$
implying that the system remains disordered.
If instead $\lambda>1/(F-1)$ the only possible attractive fixed point is 
$n_F=0$, which will attract the solution more slowly than exponentially.
In this case the system exhibits a power-law convergence to an ordered state.
Whether this transition actually takes place and for which value of the
parameter of the initial distribution, it depends on the precise value
of $\lambda$, that is plotted in Fig.~\ref{Fig2}.

For uniform initial distribution, since $\lambda=1/(Q-1)$
it is immediate to see that a phase transition occurs
for
\be
Q_c = F
\ee
so that for $Q<Q_c$ the system is ordered, while the system remains
asymptotically disordered for $Q>Q_c$.

For Poisson initial distribution and $F>2$,
the scenario is the same, with an ordered phase for small $q$ separated
by a sharp transition from an ordered phase for large $q$.
For $F=2$ all finite values of $q$ imply a positive eigenvalue $\mu$
and hence a disordered state. Only for $q=0$ order is recovered, but of
a very trivial type, since $q=0$ means that all variables are the same
{\em already in the initial state}.
Notice that, since $\lambda$ decays as $q^{-1/2}$ for large $q$, the
transition point diverges for large $F$ as $q_c \sim F^2$, differently from the
uniform case.

\begin{figure}
\centerline{\psfig{figure=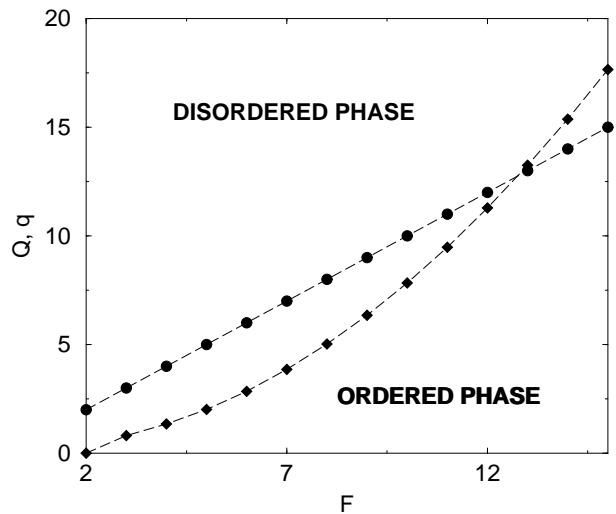,width=7cm,angle=-90}}
\caption{Mean-field phase diagram of the model. The two critical lines
are for uniform initial distribution (circles) and for Poisson initial
distribution (diamonds).}
\label{Fig4}
\end{figure}

For easier insight into the behavior of the model we plot in
Fig.~\ref{Fig4} the mean-field phase diagram.

\subsection{Uniform initial distribution with $F=2$}

Let us now give a closer look at the case of uniform initial distribution
with $F=2$.
In this case the set of mean-field equations~(\ref{MFeq}) is reduced
to two coupled equations
\bea
\dot{n}_1 &=& -\left[{3 \over 4} + {1 \over 4(Q-1)}\right] n_1^2 +
{2-Q \over 2(Q-1)} n_1 n_F \\
\dot{n}_F &=& {1 \over 4} n_1^2 - {1 \over 2(Q-1)} n_1 n_F
\label{MFF=2}
\eea
where $n_0$ has been eliminated using normalization.
In the case $Q=2$ it is easy to solve~(\ref{MFF=2}) explicitly,
since the equation for $n_1$ does not
depend on $n_F$. One obtains
\bea
n_1(t) &=& {1 \over 2+t} \\
n_F(t) &=& {1 \over (2+t)^{1/2}} - {1 \over 2 (2+t)}.
\label{MFQ=2}
\eea
Hence we find, as expected, that both the density of active bonds and of
fixed bonds vanish asymptotically as power-laws.

For $Q>2$ the two equations~(\ref{MFF=2}) become fully coupled.
The asymptotic value of $n_F$ is found to be
\be
n_F(t=\infty) = {Q-2 \over Q}
\label{nFinfty}
\ee
for all values of $Q$, which will be shown below to agree very well
with numerical results.

\section{Numerical results}
\label{Sec3}

In order to validate the results of the theoretical approach we compare
them with extensive numerical simulations of the model.
For each value of the parameters, we average over 10000 or more realizations.
For both types of initial distributions, all values of $F$ and of $q$ or $Q$,
we find that the system reaches for long times an absorbing state with no
active bonds ($n_A=0$). Extrapolation to the limit $L \to \infty$ indicates
that also in the thermodynamic limit the system freezes
asymptotically. This is the first prediction of the mean-field approach that
is verified numerically. We now turn to a more detailed analysis of the
asymptotic state as parameters are changed.

\subsection{Uniform initial distribution}

\begin{figure}
\centerline{\psfig{figure=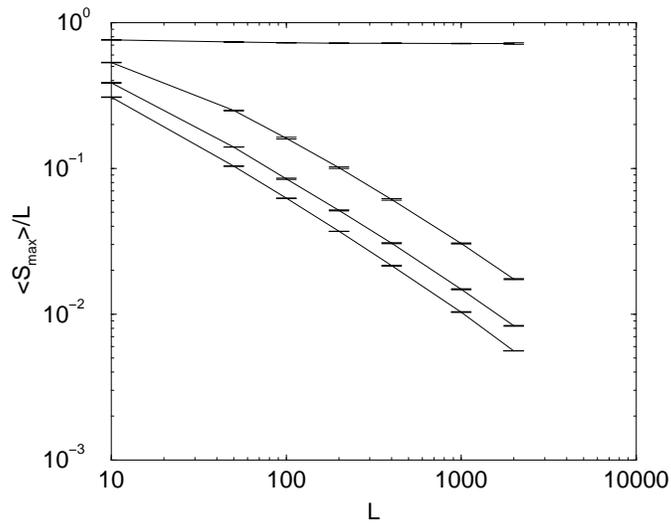,width=7cm,angle=-90}}
\caption{The order parameter
$\langle S_{max} \rangle/L$ as a function of the system size $L$ for $F=2$.
From top to bottom the curves are for $Q=2$, $Q=3$, $Q=4$ and $Q=5$.}
\label{Fig5}
\end{figure}

Let us consider first the case with $F=2$ variables on each site, which
in two dimensions gives a continuous transition.
The mean-field approach predicts the existence of a transition
between an ordered and a disordered phase for $Q_c=2$.
Such a prediction is confirmed by the behavior of the order
parameter $\langle S_{max}\rangle/L$.
As Fig.~\ref{Fig5} illustrates, $\langle S_{max}\rangle/L$ keeps for
$Q=2$ a finite value close to 0.71 in the thermodynamic limit,
while it goes to zero for $Q>2$.
Hence one macroscopic region occupies a finite fraction of the
system for $Q=2$, while for larger values of $Q$, no macroscopic
domain is formed.

We remark that in the ordered case the asymptotic value of 
$\langle S_{max}\rangle /L$ is not 1.
This should not be taken as evidence that the
system reaches a partially ordered state with the largest region
spanning only a finite part of the whole lattice.
In fact the situation is more complex.
$\langle S_{max} \rangle $ is the average over a large number of
realizations. In many of them the system becomes fully ordered and
$S_{max}/L=1$. However in a finite fraction of the realizations the
system remains partially disordered so that the order parameter
has a value smaller than 1. In practice the distribution of the
values of $S_{max}/L$ for $L \to \infty$ is a $\delta$-function for
$S_{max}/L = 1$ plus a finite nontrivial contribution for $S_{max}/L<1$
This leads to $\langle S_{max} \rangle /L \simeq 0.71$.

\begin{figure}
\centerline{\psfig{figure=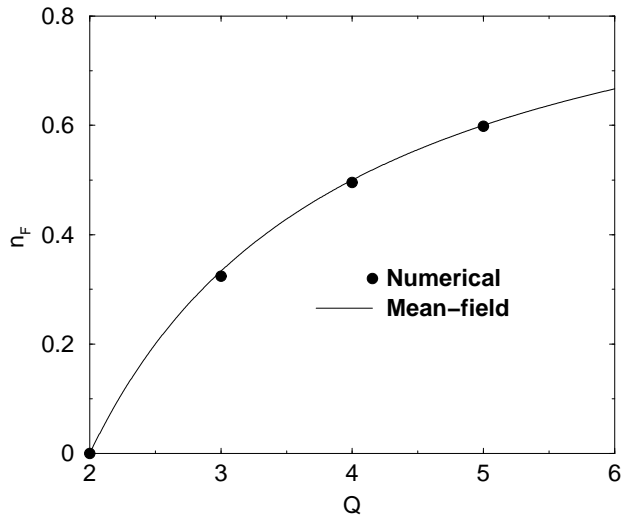,width=7cm,angle=-90}}
\caption{Comparison of the numerical results with the mean-field formula
$(Q-2)/Q$ for the final density of fixed bonds in the case $F=2$.}
\label{Fig6}
\end{figure}

For $Q>2$ the system remains disordered and a finite fraction $n_F$ of
the bonds is fixed. In the previous Section, Eq.~(\ref{nFinfty})
is the mean-field result for $n_F$ in the asymptotic state as a function
of $Q$. Such a formula is compared with numerical results in
Fig.~\ref{Fig6}.
The agreement is excellent, indicating that the mean-field approach
is {\em quantitatively} correct in this case.

\begin{figure}
\centerline{\psfig{figure=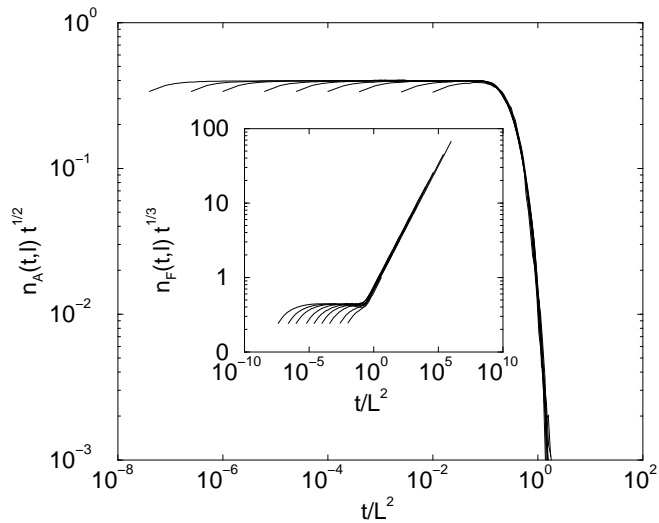,width=7cm,angle=-90}}
\caption{Main: Scaling plot of the density of active bonds $n_A$ versus time
for $F=2$, $Q=2$ and system size (left to right)
$L=10, 20, 50, 100, 200, 400, 1000, 2000, 5000$.
Inset: Scaling plot of the
density of fixed bonds $n_F$ versus time for the same set of parameters.}
\label{Fig7}
\end{figure}

Let us now analyze the whole dynamical evolution.
In the main part of Fig.~\ref{Fig7} we plot the temporal behavior of
the density of active bonds $n_A$ in the case $Q=2$, for different
system sizes.
All curves collapse on the same one when time is rescaled by $L^2$,
i. e. $n_A$ obeys the scaling form
\be
n_A(t,L) = t^{-1/2} g_A(t/L^2).
\ee
The scaling function $g_A(x)$ goes to a constant as $x \ll 1$ and goes to
zero exponentially for $x \gg 1$.
In the inset of the same figure it is shown that also the density of
fixed bonds obeys a scaling form
\be
n_F(t,L)= t^{-1/3} g_F(t/L^2)
\ee
where $g_F(x) = \mbox{const}$ for $x \ll 1$ and $g_F(x) \sim x^{1/3}$
for $x \gg 1$.

The physics behind Fig.~\ref{Fig7} is clear.
In the initial state a finite number of active bonds is present.
Notice that for $F=2$ and $Q=2$ a single type of active bond exists,
connecting sites with one equal and one different feature.
Interaction between sites implies diffusion of active bonds
and annihilation whenever two active bonds meet.
Fixed bonds are affected by active bonds but the converse is not true:
As far as active bonds are concerned, the system is equivalent to
the reaction-diffusion system $A+A \to 0$.
The decay as $t^{-1/2}$ of the density of active bonds is easily related
to the combination of diffusive wandering and mutual
annihilation~\cite{Redner}.
The inverse of $n_A$ is a correlation length that grows in time.
In an infinite system such length grows indefinitely. In a finite
lattice, growth stops when $1/n_A$ reaches the system size $L$, and
this yields a cutoff time scaling as $L^2$.
The evolution of fixed bonds is enslaved by the dynamics active bonds
but it is slower.
$1/n_F \sim t^{1/3}$ is a correlation length non equivalent to $1/n_A \sim
t^{1/2}$.
It grows with time but remains finite when the cutoff time is reached,
so that the in the frozen state $n_F$ reaches a finite value decreasing
as $L^{-2/3}$, which vanishes in the thermodynamic limit.

A comparison of the exponents found numerically with the mean-field values
[Eq.~(\ref{MFQ=2})] indicates that the mean-field approximation
correctly reproduces that $n_F$ decays more slowly than $n_A$, but it
does not predict the correct exponents.
This is not a surprising result: we have computed the mean-field rates
assuming the system to be homogeneous, while instead ordered regions grow
indefinitely.
The problem is here the same of the mean-field treatment for
the reaction-diffusion system $A+A \to 0$~\cite{Redner}.
For higher values of $Q$ the disagreement between the mean-field and
numerical results is similar.
It is somewhat striking that despite the dynamics is not captured
quantitatively by the mean-field approximation, the asymptotic value
for $n_F$ is remarkably accurate (Fig.~\ref{Fig6}).

\begin{figure}
\centerline{\psfig{figure=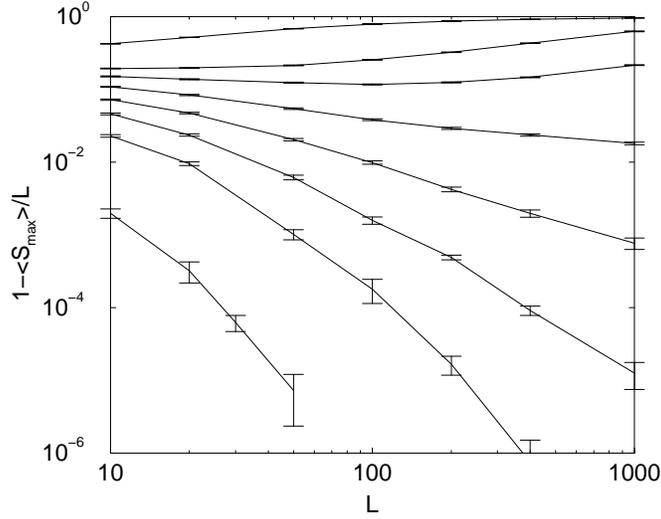,width=7cm,angle=-90}}
\caption{Plot of
$1-\langle S_{max} \rangle/L$ as a function of the system size $L$ for $F=10$.
From bottom to top the curves are for $Q=3$, $Q=5$, $Q=6$, $Q=7$, $Q=8$, $Q=9$,
$Q=10$ and $Q=15$.}
\label{Fig8}
\end{figure}

Let us now discuss the case $F>2$ corresponding to a
discontinuous transition in $d=2$.
For illustrating the behavior common to all values $F>2$ let us
focus on the case $F=10$.
As in the two-dimensional case~\cite{Castellano00}, the system exhibits
a discontinuous transition between a fully ordered phase for small $Q$
and a disordered phase for $Q>Q_c$.
This can be inferred from Fig.~\ref{Fig8}, where
$1-\langle S_{max} \rangle /L$ is plotted versus $Q$.
By observing that the convergence of $\langle S_{max} \rangle/L$ to 1
is exponential for $Q \leq 6$, while curves are bent upward for $Q>7$ one
is lead to conclude that the transition occurs for $Q_c \simeq 7$.

\subsection{Poissonian initial distribution}

Let us consider now a system such that the initial values of the variables
are chosen according to a discrete Poisson distribution of parameter $q$.
$q$ plays here a role similar to the parameter $Q$ for the uniform initial
distribution, since it fixes the average value of the variables in the
initial state.
However, no matter how small is $q$, arbitrarily large values can
be extracted with a Poisson distribution.

\begin{figure}
\centerline{\psfig{figure=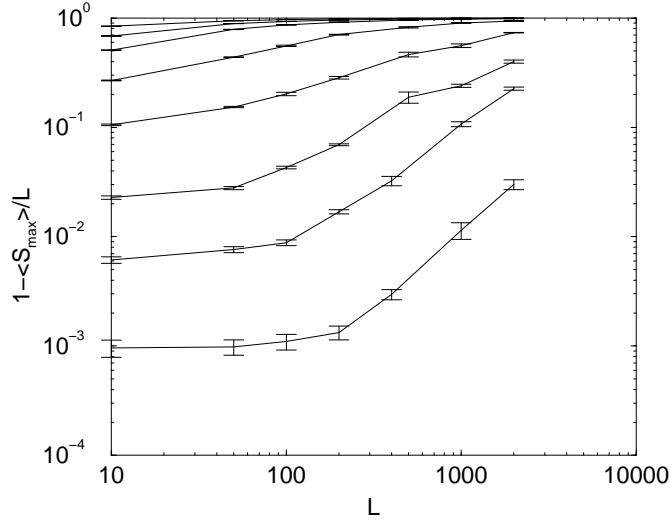,width=7cm,angle=-90}}
\caption{Plot of $1-\langle S_{max} \rangle/L$ as a function of the system
size $L$ for $F=2$ for Poisson initial distribution.
From bottom to top the curves are for $q=0.02$, $q=0.05$, $q=0.1$, $q=0.25$,
$q=0.5$, $q=1$, $q=2$ and $q=10$.}
\label{Fig9}
\end{figure}

Let us consider first the case $F=2$ (Fig.~\ref{Fig9}).
The behavior differs qualitatively from the uniform case.
For any value $q>0$ it is clear that the order parameter goes
asymptotically to zero, even if for small $q$, $\langle S_{max} \rangle /L$
remains close to 1 for relatively large $L$.
We conclude that there is strong evidence that for $F=2$ an
initial distribution of Poisson type destroys long range
order for any value of $q>0$, as predicted by the mean-field approach.

\begin{figure}
\centerline{\psfig{figure=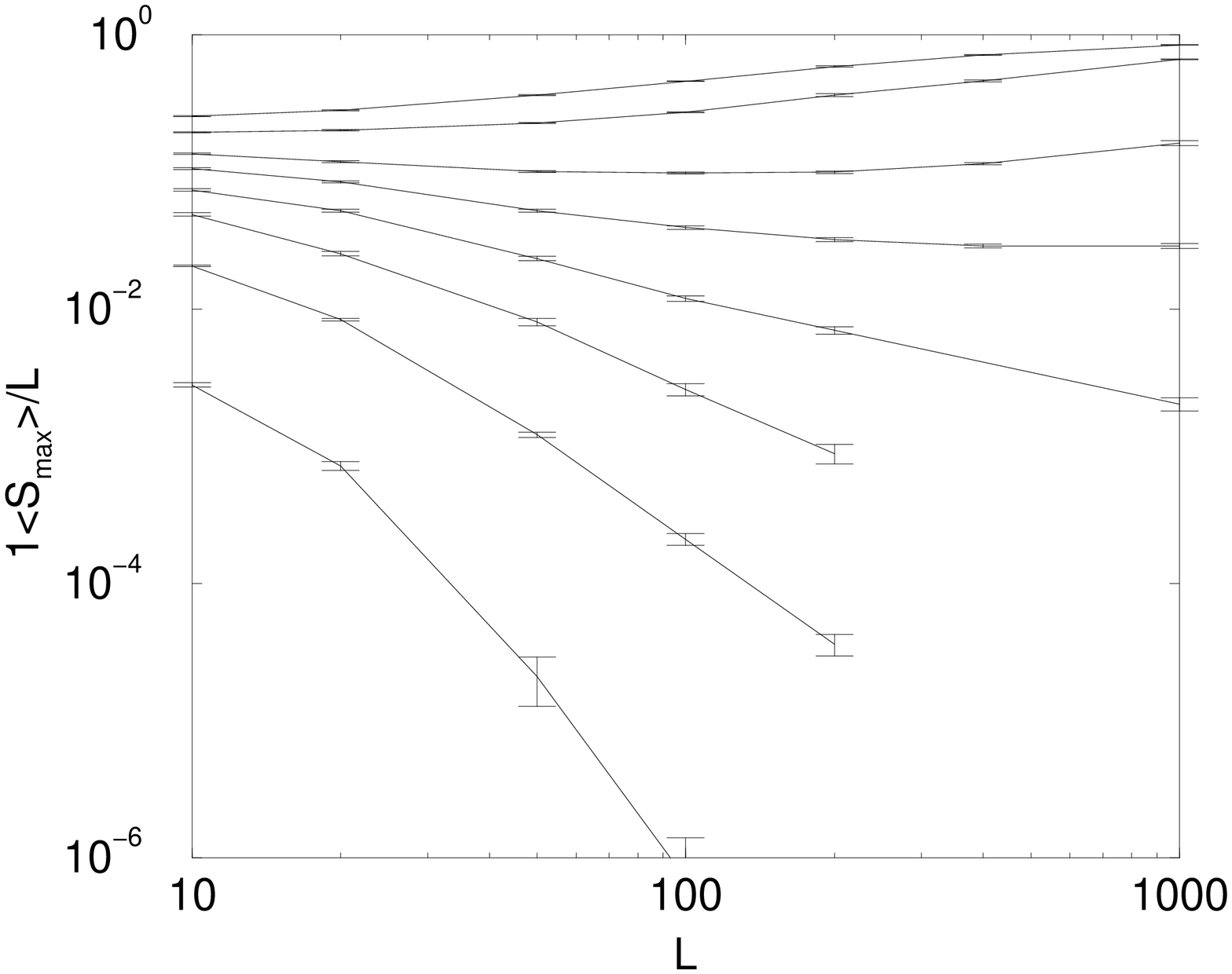,width=7cm,angle=-90}}
\caption{Plot of $1-\langle S_{max} \rangle/L$ as a function of the system
size $L$ for $F=10$ for Poisson initial distribution.
From bottom to top the curves are for $q=1$, $q=2$, $q=3$, $q=4$,
$q=5$, $q=6$, $q=8$, and $q=10$.}
\label{Fig10}
\end{figure}

We finally turn to the case $F=10$, as an example of what happens in general
for $F>2$.
The value of $1-\langle S_{max} \rangle /L$ is plotted in
Fig.~\ref{Fig10}.
For small $q$ it is clear that the system becomes ordered 
as $L$ grows while for large $q$, the curves tend to go toward 1, 
indicating the existence of a discontinuous phase transition for
intermediate $q$.
A rough estimate of the transition point is between $q=4$ and $q=5$.

These numerical data corroborate the picture emerging from the
mean-field approach.
For $F=2$ we actually find no evidence of long range order in the system,
no matter how small $q$ is taken.
For $F=10$ we observe instead the pattern already found with uniform
initial distribution: a transition between a disordered
phase for large $q$ and an ordered one for smaller values of the parameter
$q$. Also the location of the transition seems to be captured with good
accuracy.

\section{Conclusions}
\label{Sec5}
In this paper we have analyzed in detail the behavior in one dimension
of the Axelrod model, a cellular automaton introduced in Ref.~\cite{Axelrod97}
for the description of social influence.
Interestingly, the one-dimensional version of the model displays a
phenomenology as rich as the two-dimensional case. More precisely, we find
by numerical simulations a non-equilibrium phase transition
between a disordered and an ordered phase.
This phenomenology is recovered accurately by a simple mean-field
treatment of the problem, which exploits a mapping onto a reaction-diffusion
problem involving $F$ different species.
The mean-field approach is successful in several aspects. First, it correctly
predicts that the system always reaches an asymptotically absorbing state.
Secondly, it captures the existence of a transition in the final absorbing
state or the absence of it (for Poissonian initial conditions and $F=2$).
Finally, it provides
reasonably accurate estimates of the critical value of the control parameter
$Q$ or $q$. Mean-field fails instead in the prediction of the exponents
of the temporal behaviors close to the transition.
This is a typical scenario for mean-field treatments: transition points
can be computed with good accuracy while exponents are not captured correctly.
The theoretical approach presented in this paper is the simplest suitable
for this kind of problem.
For more precise values of the exponents one should go beyond mean-field,
by including diffusion terms and noise in Eq.~(\ref{MFeq}) and applying
dynamic renormalization group techniques~\cite{Peliti85,Lee95}.


\begin{thebibliography}{99}
\bibitem{Marrobook}
J. Marro and R. Dickman,
{\em Nonequilibrium Phase Transitions in Lattice Models},
Cambridge University Press, (Cambridge, U. K., 1999).

\bibitem{Jensenbook}
H. J. Jensen, 
{\em Self-Organized Criticality}
(Cambridge University Press, New York, 1998).

\bibitem{Axelrodbook}
R. Axelrod,
{\em The complexity of cooperation}
(Princeton University Press, Princeton, 1997).

\bibitem{Axelrod97}
R. Axelrod,
Journal of Conflict Resolution,
{\bf 41}, 203 (1997).

\bibitem{Axtell96}
R. Axtell, R. Axelrod, J. Epstein and M. D. Cohen, 
Computational and Mathematical Organization Theory, 
{\bf 1}, 123 (1996).

\bibitem{Castellano00}
C. Castellano, M. Marsili, and A. Vespignani,
Phys. Rev. Lett. {\bf 85}, 3536 (2000).

\bibitem{Klemm02}
K. Klemm, V. M. Eguiluz, R. Toral, and M. San Miguel,
cond-mat/0205188 (2002).

\bibitem{Liggett85}
T. M. Liggett, {\em Interacting Particle Systems}
(Springer, New York, 1985).

\bibitem{Dornic01}
I. Dornic, H. Chat\'e, J. Chave and H. Hinrichsen
Phys. Rev. Lett. {\bf 87}, 045701 (2001).

\bibitem{Wattsbook}
D. J. Watts,
{\em Small Worlds: The Dynamics of Networks between Order and Randomness},
(Princeton University Press, Princeton, 1999).

\bibitem{Strogatz01}
S. H. Strogatz, Nature {\bf 410}, 268 (2001).

\bibitem{note}
We have not been able to proof generically for any $F$ that $F-2$
eigenvalues have negative real parts and that $\mu$
is real and changes sign for $\lambda=1/(F-1)$.
However the validity of these statements can be verified by explicitly
computing the eigenvalues for any $F$.

\bibitem{Redner}
S. Redner in
{\em Nonequilibrium Statistical Mechanics in One Dimension}, 
edited by V. Privman (Cambridge,  U. K., 1997).

\bibitem{Peliti85}
L. Peliti,
J. Physique {\bf 46}, 1469 (1985).

\bibitem{Lee95}
B. P. Lee, and J. Cardy,
J. Stat. Phys. A {\bf 80}, 971 (1995).

\end{thebibliography}
\end{document}